\documentclass{icst}

\usepackage{moreverb}

\usepackage[breaklinks,colorlinks,bookmarksopen,bookmarksnumbered,linkcolor=ICSTblue,citecolor=blue,urlcolor=ICSTblue]{hyperref}
\usepackage{breakurl}
\usepackage{doi}
\usepackage{tikz}
\usetikzlibrary{shapes,arrows,calc}
\usepackage{listings}
\usepackage{graphicx}
\usepackage{adjustbox}
\usepackage{csquotes}
\usepackage[T1]{fontenc}

\newcommand\BibTeX{{\rmfamily B\kern-.05em \textsc{i\kern-.025em b}\kern-.08em
T\kern-.1667em\lower.7ex\hbox{E}\kern-.125emX}}

\def\volumeyear{2014}

\journalname{Scalable Information Systems}
\articletype{Research Article}
 \def\volumeyear{2018}
 \def\volumenumber{5}
  \def\issuemonth{04 2018 - 05 2018}
  \def\issuenumber{17}
  \def\articleid{e4}
  \def\copyrightholder{Sarathkumar Rangarajan, licensed to ICST}
  \copyrightnote{This is an open access article distributed under the terms of the Creative Commons Attribution license (\url{http://creativecommons.org/licenses/by/3.0/}), which permits unlimited use, distribution and reproduction in any medium so long as the original work is properly cited.}
  \def\articledoi{10.4108/eai.29-5-2018.154809}
\received{05 April 2018}
  \accepted{05 April 2018}
  \published{29 May 2018}

\begin{document}

\runningheads{S.Rangarajan}{Qos-Based Web Service Discovery And Selection Using Machine Learning}

\title{Qos-Based Web Service Discovery And Selection Using Machine Learning}

\author{Sarathkumar Rangarajan}

\address{Centre for Applied Informatics, Victoria University, Melbourne, Australia, VIC-3011}

\abstract{In service computing, the same target functions can be achieved by multiple Web services from different providers. Due to the functional similarities, the client needs to consider the non-functional criteria. However, Quality of Service provided by the developers suffers scarcity and lack of reliability. In addition, the reputation of the service providers is an important factor, especially those with little experience, to select a service. Most of the previous studies were focused on the user's feedbacks for justifying the selection. Unfortunately, not all the users provide the feedback unless they had extremely good or bad experience with the service. In this vision paper, we propose a novel architecture for the web service discovery and selection. The core component is a machine learning based methodology to predict the QoS properties using source code metrics. The credibility value and previous usage count are used to determine the reputation of the service.}


\fnotetext[1]{Corresponding author.  Email: \email{sarathkumar.rangarajan@live.vu.edu.au}}

\maketitle

\section{Introduction}

Due to its features including mobility, flexibility, governance, compliance, collaboration and security, SOA is the order of Internet of Things(IoT) era~\cite{DBLP:journals/corr/QinSFDWV14}. The Information Technology organizations needs to respond to this ever-changing requirements~\cite{7350251,zhang2016secure}.

Web Service(WS) is a vital technology for implementing SOA to provide interoperability between heterogeneous systems and integrating inter-organization applications\cite{wang2003achieving}. WS are Flexible, self-defining, and loosely coupled prgrammed codes running on a remote server that can be publicised, discovered, and invoked across by using a set of standard protocols through internet \cite{d2008qos}. Web Service Description Language (WSDL) is a syntactic based way to describe them.

There are quite a number of research going on efficient discovery and selection of web services. Finding the best suitable Web Service as per the user's requirements is still an open problem. Enormous numbers of operationally similar web services' availability demands non-functionally properties as a criterion in the service discovery. Quality of Services (QoS) such as execution time, cost, reliability, availability, etc., could be the valid non-functional criteria for the efficient discovery. There are many types of research focusing on QoS-based service selection but obtaining the QoS properties is a challenging task. Moreover, the QoS data are inclined significantly by various external factors such as user's geographical locations Internet connection speed between users and Web Services. Therefore, the same Web Service can provide different QoS for the different user. Added to that, some mendacious service providers may provide false QoS details. 

Moreover, if more than one services match all the criteria of the request, then the inexperienced users look for some reputation factors to choose the best service. Existing methods only depend on the previous user's feedback to justify the reputation of the services~\cite{wang2015special}. However, it is possible for the service providers to fake the feedbacks for increasing their business. Moreover, users will provide the feedback about the service only when they are very much satisfied or extremely unsatisfied. 

\subsection{Motivation to our research}
Motivation for this research is to examine the connection between QoS characteristics (Ex. reaction time, accessibility, throughput, testability, interoperability, etc.) and source code metrics (Ex. Coupling between object classes, Lack of cohesion in methods).
 
The following previous research works unearthed the possibilities to achieve the goal of this research:

\begin{itemize}
	\item The metric suite proposed by Chidamber and Kemerer comprises Weighted methods per class, Depth of Inheritance Tree,  Number of Children, Coupling between object classes, Response for a Class, Lack of cohesion in methods\cite{chidamber1994metrics}.  
\end{itemize}
\begin{itemize}
	\item The complexity, quantity and quality metrics for web service interfaces were implemented using sneed's tool.\cite{sneed2010measuring}. 
Baski and Misra's metric suite used to evaluate XML web service quality. Their proposed suites contains data weight of web service description language, message entropy metric, distinct message count and message repetition scale metric\cite{baski2011metrics}.
\end{itemize}
\begin{itemize}
	\item Coscia et al. introduced a statistical correlation analysis to predict the quality properties of WSDL documents using the classic software engineering metrics~\cite{coscia2012predicting}.
\end{itemize}
\begin{itemize}
	\item Kumar et al. utilized Principal Component Analysis (PCA) and Rough Set Analysis (RSA) as a data pre-processing process to extract and select the features to find out the correlation between 15 QoS parameters and 37 source code measurements\cite{kumar2017maintainability}. They compared three sets of metric suites and found out that the Sneed’s metric suite provides superior results over all the three. 
\end{itemize}

We planned to use linear regression to develop the training set according to the correlation standards defined in \cite{kumar2017maintainability}. Then the number of latent variables need to be defined for each QoS property. The next step is to build a model to generate the value of the QoS property by using the training set knowledge base.
For this research, we used QWS-WSDLs Dataset Version 2.0 for the WSDL list as well as the QoS properties of the Web Services~\cite{al2007qos}. QWS Dataset Version 2.0 includes a set of 2,507 Web Services and their QWS measurements that were conducted in March 2008 using our Web Service Broker (WSB) framework. We requested the owner of the data set to utilize for this research and the Al-Masri et al. are happy to provide access to the data set. Even though QWS data set is being used as a bench mark data set in the recent researches, we planned to validate the data set by checking the link to its WSDL. The data set is provided with the URL for the WSDL file. The URL can be used to check the availability of the WSDL. Once the validation is completed, the data set will be divided into two data sets such as part 1 and part 2. The part 1 data set will be used to train the learning machine and part 2 will be used as test set to validate the accuracy of the prediction. Mean Absolute Error (MAE) and Root Mean Squared Error (RMSE), the most used statistical accuracy metrics will be used to evaluate the proposed methods' prediction accuracy.

Hence, there is a need to develop a novel architecture to discover and select WSDL-based Web Services using QoS properties. To develop an efficient architecture, the proposed research should answer the following research questions:

\textbf{\textit{RQ1 How to extract QoS properties for Web Service without provider's contribution?}}\\
A noteworthy issue in utilizing QoS for service discovery is the instability of the QoS data and lack of sufficient parameters. The greater part of the QoS-aware discovery components depicted in the past research techniques, disregard this issue. To gather adequate Web Service QoS information, the current frameworks requires assessments from various geographic areas under different system conditions. On the other hand, it is not a simple task to do a vast scale dispersed web services assessment. In this research, we endeavour to bridge this research gap by using learning machines to predict QoS properties in using source code metrics of the service. 

\textbf{\textit{RQ2 How to justify the reputation of the service providers?}}\\
With more and more Web Services being deployed on the Web, service selection methods may provide a pool of proper Web Services. All things considered, customers have more contrasting options to choose. Some of the purchasers may have next to zero knowledge of the service and services at the cold start state leads to complexity in decision-making. Reputation systems were introduced to moderate the hazard that purchasers confronted while choosing a service. The current reputation frameworks mostly based on the response from the past clients. Unless the service is too good or too bad most of the consumers are not interested in providing the rating about their transaction. In this study, we make use of the calculated QoS properties and compare them with the provider's QoS properties to find the credibility of the provider. Along with the credibility factor, previous successful usage count of the service will also be accounted to justify the reputation of the service to help the consumer in WS selection.

\subsection{Contributions}
In this study, we particularly aim to develop an effective response to the challenges currently faced by web service discovery and selection. We focus on the development of a novel architecture designed to support web service consumers in their discovery and selection of the WS. 

The contributions of this research are:

\begin{itemize}
	\item A novel architecture to improve the discovery and selection of the WS (Section~\ref{sec:archi}).
	\item Machine learning based technique to predict Quality of Service (QoS) from source code metrics (Section~\ref{sec:MLpredi}).
	\item The reputation estimation method using credibility of WS provider and usage history of WS (Section~\ref{sec:repu}). 
\end{itemize}
Service consumers search and review the description of the services offered by service providers and select a service~\cite{wang2002ticket}. Existing service reputation systems, mainly based on the ratings given by service consumers, are one of the most important guides that the consumer has in deciding, as they reveal how other consumers evaluated the services true ability in real scenarios. But, these works were suffered by skewed ratings by the consumers and cold-start stages. Hence this research identifies the credibility of the service provider by comparing the predicted QoS and developer assured QoS.
\subsection{Significance of the research}
A developing issue faced by IT administrators will be how to coordinate remote applications and versatile wireless devices under corporate frameworks~\cite{wang2006ubiquitous,khalil2009integrated,khalil2007integrating}. Web service give an approach about coordinating remote applications and mobile devices; without blowing the IT infrastructure budget plan on complex and expensive collaborative projects~\cite{Li2011}. Certainly, there are 50,000 programmers signed up for Amazon Web Services program, and 8,000 organizations and individual programmers appreciating membership in eBay API program.  Due to the overabundance of services, there are a bunch of functionally alike web services are presented in registries. Consequently, the WS discovery and negotiation demands an quality based criterian to choose a valid service for the business~\cite{wang2005flexible}.

The propagation of Web Services into our businesses and day-to-day lives has made the quality of services (QoS) a very important aspect for both the service provider and the consumers. With many Web Service providers providing similar services, Web Service selection based on QoS classification becomes crucial for the consumer.

To the best of my knowledge, this would be the first work to use the correlation between object oriented programming source code metrics and QoS attributes to find the QoS properties for the Web Services. This research automates the process of identifying the QoS to enhance the wireless application transactions.

\section{Proposed architecture for web service discovery and selection}
\label{sec:archi}

To address the challenges in the web services selection and ranking problem, we propose an architecture, which mainly consists of the following modules:
\begin{itemize}
	\item Standardisation of WS description
	\item Web Service Discovery
	\item Machine learning for QoS prediction
	\item Web Service Selection
	\item Reputation estimation
\end{itemize}

\begin{figure}
	\centering
	\includegraphics[width=0.9\linewidth]{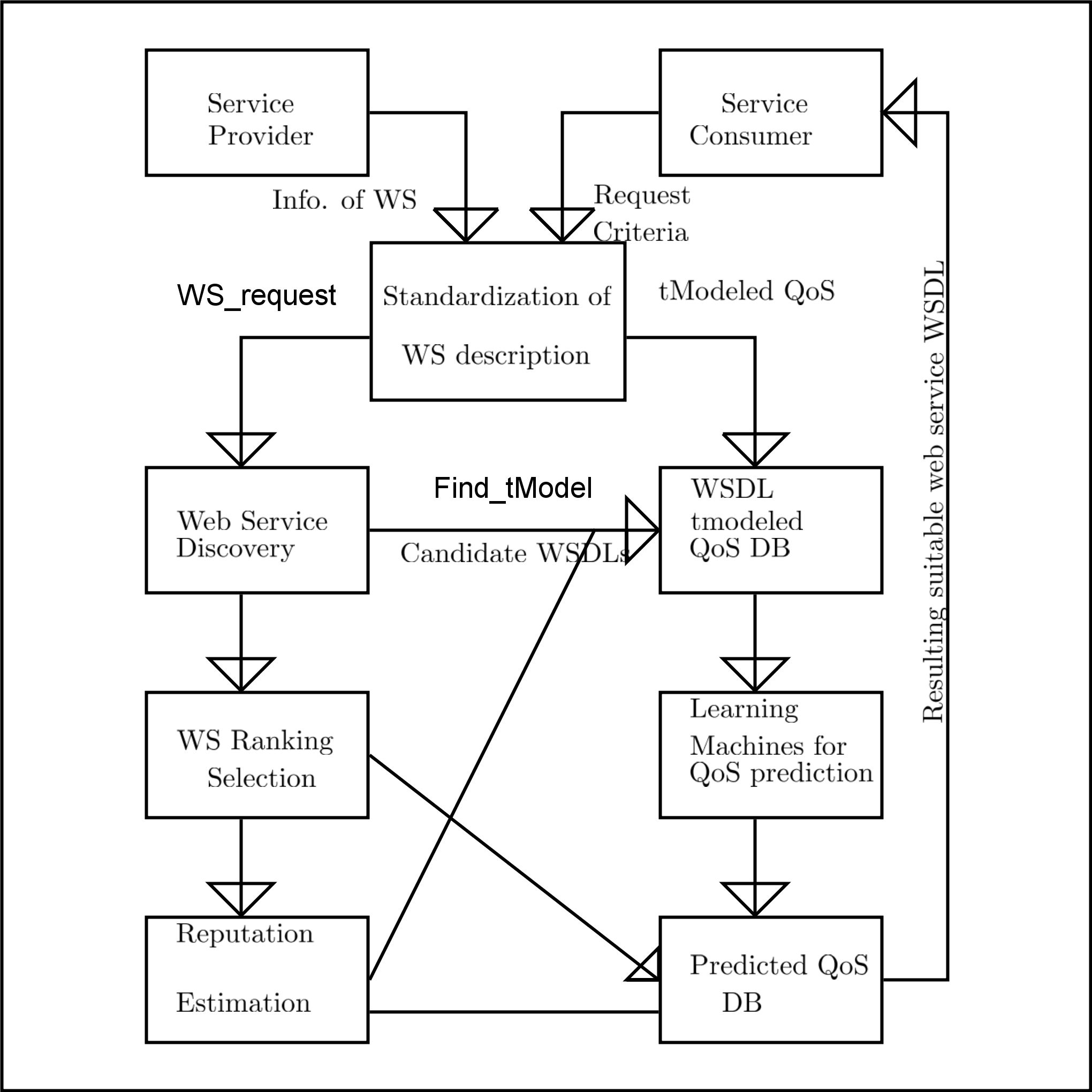}
	\caption[WS archi]{Architecture for Web service discovery and selection}
	\label{fig:wsarchi}
\end{figure}

Figure~\ref{fig:wsarchi} shows the architecture for the proposed web service discovery and selection method. Standardization of the WS description has two parts. First, the QoS properties provided by the service providers for a new service needs to be converted into a XML based data structure named tModel. The second part is to formulate the service request information as service request message. 

On the arrival of new tModel based QoS information and WSDL, the learning machine utilise the WSDL file to extract the class files and identify the source code metrics values for the class files. Learning machines are used to identify the QoS properties for the metrics using the training data. The predicted QoS data will be stored in the predicted QoS database with unique key to identify during the selection process. 

On the other hand, the module for web service discovery will handle a new service request. The candidate WSDLs retrieved from the database using the find\_tModel method. Selection of the target service is the next step to be done. Web service ranking module makes use of the predicted QoS for the candidate WSDLs and quality constrain tree to rank the services against the request parameters. Reputation estimation needs predicted QoS properties, service provider's QoS properties and the usage history of the services. It calculates the reputation score for the services and the service with the maximum rank will be considered as the resulting service. 

In the rest of this section, we will present the technical details for the modules of standardisation of WS description, WS discovery, and WS selection. In Section~\ref{sec:MLpredi}, we will discuss machine learning for QoS prediction. Reputation estimation will be discussed in Section~\ref{sec:repu}.

\subsection{Standardisation of the WS description}

\subsection{WS request message}
There are a large number of QoS properties available. A sub set consists of five QoS properties such as 1) Response Time, 2) Availability, 3) Throughput, 4) Reliability, 5) Latency were considered for this research. Among the other QoS properties these five properties are categorized as the key values to measure the performance of the WS \cite{wang2012discovery}.

\lstset{language=XML,morekeywords={find\_service,functionalReq,qualityReq,property,value,weight,MaxService,tModel,function,overviewDoc,categoryBag,find\_tModel,keylimit}}

\begin{adjustbox}{width=\linewidth}
\begin{lstlisting}[caption={Request message example},label={lst:req_msg}]
<find\_service generic="1.0" xmlns="urn:uddi-org:api">
 <functionalReq>credit card validation</functionalReq>
	<qualityReq>
		<property>price</property>
		<value>0.01</value>
		<weight>2</weight>
	</qualityReq>
	<qualityReq>
		<property>Response time</property>
		<value>0.05</value>
		<weight>3</weight>
	</qualityReq>
	<MaxService>2<\MaxService>
</find\_service> 
\end{lstlisting}
\end{adjustbox}

Web Service request must contain the functional and non-functional criteria for the target Web Service in standardised XML messages. Harshavardhanan et al. proposed a standard to define the client request message using XML \cite{harshavardhanan2012dynamic}. We used their message standard for defining the client's request. For each quality measures, the client can give the weight value to prioritise the QoS attributes.

For example, if the consumer needs a Web Service with the best availability property, he/she can compromise the price for it. Then he/she needs to give high weight value (High-5 to low-1) for the availability whereas the weight value of the price should be lower. Listing~\ref{lst:req_msg} shows an example XML SOAP message of a sample request message for the credit card validation Web Service request.  

\subsection{Data structure for the WS information}
The functional and QoS properties of a Web Service can be represented using the data structure called tModel proposed in \cite{rajendran2009analysis}. The role of a tModel is to register categorizations, which provides an extensible mechanism for adding property information to a Web Service registry. It is used to provide QoS information on bindingTemplates. For each QoS parameter, a general-purpose name-value pair structure is used to represent the QoS property name and value. The example showed in listing~\ref{lst:tmod_ex} is for the QoS details of a Stock Quote service. A tModel with tModelKey "mycompany.com:StockQuoteService: PrimaryBinding: QoSInformation" containing the QoS attribute categories is referenced in the bindingTemplate. To retrieve more detailed management information, the location of a WSDL description is stored in a keyed reference with tModelKey \enquote{uddi: mycompany.com: StockQuoteService}. function tag refers  the opertaion of the web Service and ws\_id tag is the unique identifier of the WS in the registry.

\begin{adjustbox}{width=\linewidth}
\begin{lstlisting}[caption={tmodel example},label={lst:tmod_ex}]
<tModel tModelKey="mycompany.com:StockQuote:QoS">
<function>Stock_Quote_Service</function>
<ws\_id>abdc12345<ws\_id>
<overviewDoc>     
<overviewURL>http://<URL of QoS schema></overviewURL>
</overviewDoc>
<categoryBag> 
 <keyedReference 
  tModelKey="uddi:uddi.org:QoS:Availability"
  keyName="Availability"
  keyValue="99.9\%"/> 
 <keyedReference 
  tModelKey="uddi:uddi.org:QoS:Throughput"
  keyName="Average_Throughput" 
  keyValue=">10Mbps"/>
 <keyedReference 
  tModelKey="uddi:uddi.org:QoS:Reliability"
  keyName="Average_Reliability"
  keyValue="99.9\%"/>
</categoryBag>
</tModel>
\end{lstlisting}
\end{adjustbox}

\subsection{Web Service Discovery}
The Service discovery for the client's request is using functional requirement detail to list the functionally similar services from the registry. All WSDL service interfaces are published in the registry as a tModel. The find\_tModel method proposed by Rajendran et al. listed all the matching tModel\_id interface descriptions \cite{rajendran2010optimal}. After a tModel\_id has been retrieved, the overviewURL can be used to retrieve the contents of the WSDL service interface document. Additional keyedReferences can be added to the categoryBag to limit the set of tModel that are returned in the response to this find request. Listing~\ref{lst:disc}, an example shows a find\_tModel message, which can be used to locate all the stock quote services available in the registry.
\begin{adjustbox}{width=\linewidth}
	\begin{lstlisting}[caption={Discovery message example}, label={lst:disc}]
<find\_tModel generic="1.0" xmlns="urn:uddi-org:api">
 <categoryBag>   
  <keyedReference tM\_find\_Key="UUID:DB77450D-9FA8" 
   keyName="Stock market trading services"
   keylimit="50"/>
 </categoryBag>
</find\_tModel>
\end{lstlisting}
\end{adjustbox}

\subsection{Web Service Selection}

In this research, we use the min-max normalisation technique and weighted AND-OR tree proposed by D'Mello et al. to rank the service based on the QoS and its weight \cite{d2008qos}. A Weighted AND-OR tree is an AND-OR tree where every edge between parent and child nodes is labeled with a non-negative real number in an interval $ (0, 1) $.For any parent node, the sum of edge labels (weights) of all child nodes is equal to one i.e. for any parent node P with C $ (2<=C<=N) $child nodes, the sum of edge weights$ WPC_i$  $ (1<=i<=C) $ is equal to one.
In $ < $conditioned leaf node:

\begin{equation}\label{eq3}
	WSS_n=\begin{cases}
		& \frac{2(WS_{last}-WS_x)}{10}\\ 
		& WSS_{first}=1 \;\;\;\;\; x= 2,3,..last-1\\ 
		& WSS_{last}=0  
	\end{cases}
\end{equation}

In $ > $ conditioned leaf node:

\begin{equation}\label{eq4}
	WSS_n=\begin{cases}
		& \frac{2(WS_{first}-WS_x)}{10}\\ 
		& WSS_{first}=0 \;\;\;\;\; x= 2,3,..last-1\\ 
		& WSS_{last}=1 
	\end{cases},
\end{equation}
where $ wss_n $ refers to the $n$th Web Service's QoS Score, $ ws_{first} $	denotes the first Web Service in the descending order in the leaf node, and $ ws_{last} $ represents the last Web Service in the descending order in the leaf node.

Equations \ref{eq3} and \ref{eq4} are used to calculate the score of the target web service. After constructing the tree the root node will have the descending ordered list of the Web Services based on their QoS Scores. In the top of the list, we can find the best Web Service based on the client's quality requirement. 

\section{Machine learning for QoS prediction}
\label{sec:MLpredi}

The bottomline of this research is to examine the connection between QoS characteristics (Ex. reaction time, accessibility, throughput, testability, interoperability, etc.) and source code metrics (Ex. Coupling between object classes, Lack of cohesion in methods). 
The following research works yields promising findings to achieve the goal of this research.
\begin{itemize}
	\item 	The metric suite proposed by Chidamber and Kemerer comprises Weighted methods per class, Depth of Inheritance Tree,  Number of Children, Coupling between object classes, Response for a Class, Lack of cohesion in methods\cite{chidamber1994metrics}.  
	\item The complexity, quantity and quality metrics for web service interfaces were implemented using sneed's tool.\cite{sneed2010measuring}. 
	\item Baski and Misra's metric suite used to evaluate XML web service quality. Their proposed suites contains data weight of web service description language, message entropy metric, distinct message count and message repetition scale metric\cite{baski2011metrics}.
	\item Coscia et al. introduced a statistical correlation analysis to predict the quality properties of WSDL documents using the classic software engineering metrics~\cite{coscia2012predicting}.
	\item Kumar et al. utilized Principal Component Analysis (PCA) and Rough Set Analysis (RSA) as a data pre-processing process to extract and select the features to find out the correlation between 15 QoS parameters and 37 source code measurements\cite{kumar2017maintainability}. They compared three sets of metric suites and found out that the Sneed's metric suite provides superior results over all the three. 
\end{itemize}
We planned to use linear regression to develop the training set according to the correlation standards defined in \cite{kumar2017maintainability}. Then the number of latent variables need to be defined for each QoS property. The next step is to build a model to generate the value of the QoS property by using the training set knowledge base.
For this research, we used QWS-WSDLs Dataset Version 2.0 for the WSDL list as well as the QoS properties of the Web Services~\cite{al2007qos}. QWS Dataset Version 2.0 includes a set of 2,507 Web Services and their QWS measurements that were conducted in March 2008 using our Web Service Broker (WSB) framework. We requested the owner of the data set to utilize for this research and the Al-Masri et al. are happy to provide access to the data set. Even though QWS data set is being used as a bench mark data set in the recent researches, we planned to validate the data set by checking the link to its WSDL. The data set is provided with the URL for the WSDL file. The URL can be used to check the availability of the WSDL. After validation, the verified web service’ data set will be divided into two groups. Group 1 will be having 80\% of the data set to act as a training set and the remaining 20\% will be used as a test data set.

Mean Absolute Error (MAE) and Root Mean Squared Error (RMSE) are the two well-known statistical precision measurements utilized to assess the prediction results. MAE is the average absolute deviation of predictions to the ground truth data. For all test services and test QoS properties, MAE is calculated as:

\begin{equation}\label{eq1}
	MAE=\frac{\left(\sum ij\left\|Q_{ij}-\hat{Q}_{ij}\right\|\right)}{N}
\end{equation}

\tikzstyle{block} = [rectangle, draw, text width=5em, text centered, rounded corners, minimum height=4em]
\tikzstyle{line} = [draw, -latex']
\tikzstyle{cloud} = [draw, ellipse, node distance=3cm,
minimum height=2em]    
\begin{figure}
	\begin{tikzpicture}[node distance = 3cm, auto]
	\node [block] (deter) {Determine no. of latent variable};
	\node [block, below of=deter, node distance=3cm] (build) {Build model};
	\node [block, below of=build, node distance=3cm] (quality) {Assign Quality value};
	\node [block, left of=build, node distance=3cm] (train) {Training set};
	\node [block, below of=train, node distance=2cm] (test) {Test set};
	\path [line] (deter) -- node {No. of latent variables}(build);
	\path [line] (build) -- node [midway, above]{Multivariant regression coefficients}(quality);
	\path [line] (train) -- (deter);
	\path [line] (train) -- (build);
	\path [line] (test) -- (quality);
	
	\end{tikzpicture}
	\caption{Learning machine flowchart}
	\label{lrmac}
\end{figure}
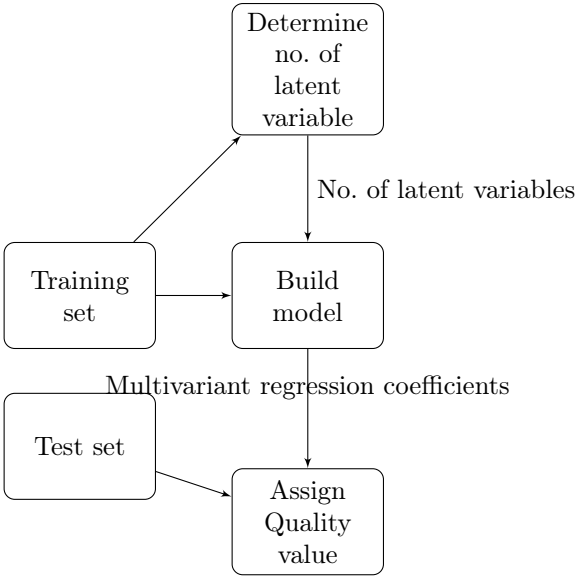

In the equation \ref{eq1}, $Q_{ij}$ denotes the observed QoS value of Web service j obtained from data set entry i; $\hat{Q}_{ij}$  is the predicted QoS value; and $N$ is the number of predicted values. The smaller value of MAE indicates better prediction result. RMSE can be expressed as:

\begin{equation}\label{eq2}
	RMSE=\sqrt{\frac{\sum ij\|Q_{ij}-\hat{Q}_{ij}\|^{2}}{N}}
\end{equation}

RMSE can be measured using the equation \ref{eq2} to find out the differences between the actual and predicted values. Once the model yields more than 90\% accuracy level, the learning machine can able to predict the five chosen QoS properties for a WSDL of a service. The predicted QoS properties stored in the predicted QoS database for the calculation of the QoS score during the selection of the service. 

\section{Reputation estimation}
\label{sec:repu}

Once the QoS properties of the Web Service is identified then it is compared with the QoS details obtained from the service providers using the credibility scoring algorithm specified in Figure~\ref{fig:wsrep}.  

The calculated $ C_{WSi} $ value will be stored in the reputation database in the registry along with the WS unique id.

This algorithm will compare each QoS parameters in the both data sets and if the identified value($ Q_{v} $) is equal to the assured value ($ Q_a $) then the credibility factor($ C_{WSi} $) will be incremented by 1 and the loop happens for all the identified 15 QoS properties and finally it will return the overall credibility value for the corresponding Web Service. Likewise, usage history also tracked for the successful transaction of the Web Service and $ WS_{count} $ value will be stored in the reputation DB. Once the WS selection algorithm ranked and returned the possible right Web Services, the reputation value will be calculated by adding $ C_{WSi} $ and  $ WS_{count} $. Finally, the Web Service that has the highest value of reputation value will be declared as the most suitable Web Service for the corresponding client request.

\tikzstyle{decision} = [diamond, draw, text width=5em, text badly centered, node distance=3cm, inner sep=0pt]
\tikzstyle{block} = [rectangle, draw,  text width=5em, text centered, rounded corners, minimum height=4em]
\tikzstyle{line} = [draw, -latex']
\begin{figure}
	\begin{tikzpicture}[node distance = 2cm, auto]
	\node [block] (qval) {$ Qv[], Qa[] $};
	\node [decision, below of=qval] (decide) {$ If Qv[i]== Qa[i] $};
	\node [block, left of=decide, node distance=3cm] (fora) {for all $ i $};
	\node [block, below of=decide, node distance=3cm] (inc) {Increment $ C_{WSi} $ by $ 1 $};
	\path [line] (qval) -- (decide);
	\path [line] (decide) -- node {no} (fora);
	\path [line] (decide) -- node {yes} (inc);
	\path [line] (inc) -| node [near start] {$i++$} (fora);
	\path [line] (fora) |- ($(qval)!.5!(decide)$);
	\end{tikzpicture}
	\caption[repu chart]{Flowchart for Credibility scoring}
	\label{fig:wsrep}
\end{figure}
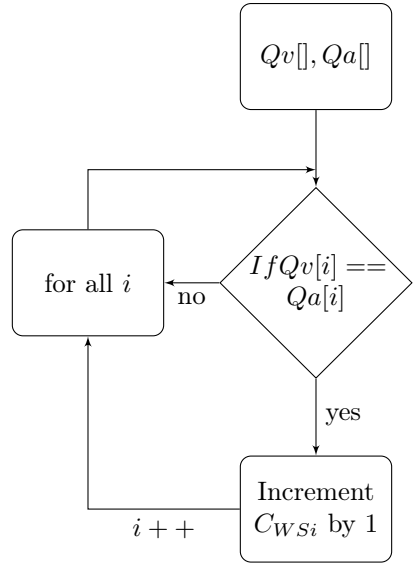

The WSDL file of the WS will be given to the client for the further transaction. Once the client initiates the connection with the Web Service the usage count variable $ WS_{count} $ will be incremented. Thus this research overcomes the less availability of user ratings and dependability of user based reviews to determine the reputation of the Web Service. The client can use the overview URL to find the details about the WSDL such as input/output parameters, types, message, portType, binding, port, service.

\section{Related works}

WSDL and UDDI are the largely utilized standards in IT businesses for describing and discovering Web Services\cite{mcilraith2001semantic}. 
The functional semantic way of describing Web Services to afford dynamic Web Service discovery was depicted by Ye and Zhang \cite{ye2006web}. Ye et al.  extended their work by describes FWSDL, Web Service description language to denote functional semantics and the structure of discovery method named FunWDS\cite{ye2006discovering}.

D.Mello et al. anticipated a well-formed functional semantics to portray operations of Web Services by giving extendible functional knowledge to map the requested or published operation descriptions into an abstract operation \cite{d2009effective}. In spite of this, the focal point lies in indicating the structure of the service interfaces and of the exchanged messages. In this way, the earlier works deal with the discovery upon problem relying on structural, keyword-based matching, which restraints their search capabilities \cite{averbakh2009exploiting}. Therefore, service clients pay additional attention to QoS as an alternative to functionality than before. QoS basically comprises of non-functional properties such as response time, throughput, availability, etc \cite{chen2015hybrid}.

Xu et al. proposed a QoS-based Web Service discovery model by expanding the data structure types to improve the UDDI model along with QoS properties \cite{xu2007reputation}. Their methodology demands the human customer to carry out the service discovery and selection. Obviously, this method is not scalable if there are a massive amount of Web Services are available as an option. Ahmed and Azam proposed a framework for selection of web services based on QoS attributes and customer's preferences set over them, and developed an algorithm that is Diversified Service Rank (DSR)\cite{ahmed2014selection}. One more method appearing in~\cite{mobedpour2013user} propose a QoS-based WS selection procedure which receives QoS request with exact values and fuzzy values and returns matching offers in both categories: super-exact and partial matches. In ~\cite{iordache2014qos}, users' preferences are characterized by a lexical ordering in accordance with their perceived importance.

Coscia et al. presented the opportunity to predict the service interface maintainability or QoS by applying conventional software metrics in service implementation. Their methodology proposed the use of Object Oriented metrics as early indicators to guide software designers in the direction of getting more maintainable services \cite{coscia2012predicting}. 

Machine learning can be divide into unsupervised, semi-supervised and supervised learnings~\cite{cao2013shilling}. Huang et al. bring up with the intention to show that the Extreme Learning Machines (ELMs) is better than the computational intelligence methods such as Artificial Neural Networks (ANNs) and Support Vector 
Machines (SVMs) in-terms of learning rate and computational scalability~\cite{huang2011extreme}. Mateos et al. establish that there is a high correlation among well-known object-oriented metrics taken in the code implementing services and the occurrences of anti-patterns in their WSDLs \cite{mateos2011detecting}. Kumar et al. used different object-oriented software metrics and Support Vector Machines with a different type of kernels for predicting maintainability of service \cite{kumar2017maintainability}.

There is a decent number of research is going on reputation models and feedback shortage problem in Web Service discovery and selection \cite{malik2009web}. Rating shortage or sparsity primarily happens at the cold-start stage of a service or when a service experiences a extended period of inactivity. The Bayesian reputation system proposed by J ~{A}.sang and Quattrociocchi \cite{josang2009advanced} addresses the significance of base rate in the cold-start stage. the Bayesian reputation system cannot boost the convergence rate.

\section{Conclusion}

Identifying a suitable web service for business entity is always an important part of the service-oriented architecture. In order to utilize the availability of non-functional criteria to choose the best suitable web services, the QoS became a decision making factor. However, the sparsity of the QoS data from the service providers and unequal environment factors had been hindrances to use the available QoS values. In this paper, we proposed a new architecture, which predicts the QoS values for the web services' source code metrics using a learning machine.

This research utilises the predicted QoS properties to find the reputation of the service providers for helping less experienced clients. The main challenge of this work to validate the dataset. The QWS dataset has 2507 WS entries and each of them needs to validated by checking its URL to locate the WSDL file. This research is limited to the WSDL based web service discovery and selection. Ontology based (OWL-S) Semantic web services cannot be discovered by using the proposed architecture.
\bibliographystyle{icstnum}
\bibliography{ref}

\end{document}